\documentstyle[12pt,aasms4,psfig]{article}
\lefthead{Ulmer et al}
\righthead{Low Surface Brightness Galaxies in Coma}

\catcode`\@=11
\def\simgreat{\mathbin{\lower 3pt\hbox
     {$\rlap{\raise 5pt\hbox{$\char'076$}}\mathchar"7218$}}} 
\def\gapprox{\mathrel{\mathpalette\@versim>}}
\def\lapprox{\mathrel{\mathpalette\@versim<}}
\def\@versim#1#2{\lower2.45pt\vbox{\baselineskip0pt\lineskip0.9pt
    \ialign{$\m@th#1\hfil##\hfil$\crcr#2\crcr\sim\crcr}}}
\catcode`\@=12
\def\keV{\rm ke\kern-0.11em V}
\def\MeV{\rm Me\kern-0.11em V}
\def\deg{$^\circ$}
\def\arcmin{\hbox{$^\prime$}}
\def\arcsec{\hbox{$^{\prime\prime}$}}

\def\ref{\noindent \hangindent 1cm }

\def\@doubleleading{1.5}
\def\baselinestretch{\@doubleleading}
\parskip=\baselineskip
\begin{document}
%
\def\@biblabel#1{\hfill}
%
%
\ \ 
\title{{\sc Low Surface Brightness Galaxies in the Core of the Coma Cluster}}

\author{M. P. Ulmer\altaffilmark{1,7}, 
G. M. Bernstein\altaffilmark{2,7}, D. R. Martin\altaffilmark{3}, 
R. C. Nichol\altaffilmark{4}, J. L. Pendleton\altaffilmark{1,5}\\ 
and J. A. Tyson\altaffilmark{6,7}}


\altaffiltext{1}{Dept.\ of Physics and Astronomy, Northwestern University, 
	Evanston, IL \ 60208-2900;
	Electronic mail:  m-ulmer2@nwu.edu}
\altaffiltext{2}{Dept.\ of Astronomy, 829 Dennison Bldg., University of
	Michigan, Ann Arbor, MI \ 48109;
	Electronic mail:  garyb@astro.lsa.umich.edu}
\altaffiltext{3}{Princeton Univ., Princeton, NJ 08544;
	Electronic mail: drmartin@phoenix.princeton.edu}
\altaffiltext{4}{Dept.\ of Astronomy and Astrophysics, 
        University of Chicago, 5640 South Ellis Ave., Chicago, IL \ 60637;
	Electronic mail:  nichol@oddjob.uchicago.edu} 
\altaffiltext{5}{Current address: Research Systems, Inc., 2995 Wilderness 
        Place, Boulder, CO \ 80301;
        Electronic mail: jpendleton@rsinc.com}
\altaffiltext{6}{AT\&T Bell Laboratories, Room 1D432, Murray Hill, NJ \ 07974;
	Electronic mail: tyson@physics.att.com}
\altaffiltext{7}{Visiting Astronomer, Kitt Peak National Observatory}

\setcounter{footnote}{0}
 

%
%
\begin{abstract} 


We present the results of a search for low surface brightness galaxies 
(hereafter LSBs) in the Coma cluster.  Bernstein et al. report on deep CCD
observations in R of a $\sim 7.\arcmin5 \times 7.\arcmin5$ region in the core
of the Coma cluster, and we extend this work by finding and measuring 36 LSBs
within this field.  We report both R and B$_j$ results.  The average magnitude
based on the best fit exponential to the images is 22.5 (R) and the typical
exponential scale is 1.\arcsec3.  The range of exponential scales is 0.4 to 1.2
kpc (distance modulus 34.89), and the range of central surface brightnesses is
24 to 27.4 R mag per square arcsecond.  Many of these objects are similar in
terms of scale length and central surface brightness to those found by others
in nearby clusters such as Fornax (Bothun, Impey \& Malin), 
as well as in the low
luminosity end of the dwarfs cataloged in the review of Ferguson \& Binggeli.  We
find no evidence for a dependence of color on central surface brightness or on
distance from the D galaxies or the X-ray center of Coma.  We also find that
these LSBs make a small contribution to the overall mass of the cluster.  We
discuss these results in terms of possible scenarios of LSB formation and
evolution.

\end{abstract} 

\newpage
\section{{\sc INTRODUCTION}} 

Due to its high galactic latitude, its richness, and relatively low redshift,
the Coma cluster of galaxies is an excellent place to study the effects of
cluster environment on the formation and evolution of low luminosity galaxies. 
Extensive studies of the Coma cluster have shown that it has a high 
mass-to-light ratio (about 300) and a dense ($10^{-3}$ particles/cm$^3$), hot
($10^8$ K), intra-cluster medium (cf. Sarazin 1986).  These characteristics
of the Coma cluster make its core region an extremely interesting environment
to study, and increased knowledge of the faint end of the luminosity function
will extend our understanding of the formation and evolution of galaxies in
rich clusters.  Here we report an analysis of low surface brightness galaxies
(LSBs, hereafter) which, for our sample, are effectively a subclass of dwarf
ellipticals (dEs, hereafter).  Our analysis was based on deep CCD imaging of
the Coma cluster core (Bernstein et al. 1995).  This study is of the richest
cluster environment yet reported, and the purpose of this paper is to find
fainter objects in denser cluster environments than previous reports: Fornax
 (Bothun, Impey \& Malin 1991; Bothun, Caldwell \& Schombert 1989; Phillipps et al. 1987  and references
therein), Virgo (Impey, Bothun \& Malin 1988; Binggeli \& Cameron 1991), 
and A3574 (Turner et al. 1993).  The LSBs we have found have central surface brightnesses as
faint as 27.4 per square arcsecond (R), and exponential scales as large as 1.2
kpc.  The projected density of these objects per square Mpc is about 8 times
the previously reported maximum value (Turner et al.).  A general definition
of LSBs is that the central surface brightness of the galaxy be fainter than 23
mag per square arcsecond in B (cf. Bothun et al. 1991), but there are a
wide range of values (1$\arcsec$ to 20$\arcsec$) for the exponential scale
factors and net size and brightness (Bothun et al. 1991;
Turner et al. 1993; Bothun et al. 1989; Binggeli \& Cameron).  The origin
of these extreme objects is still unknown, but as more data like ours are
accumulated, we will achieve better ideas of the formation processes of LSBs
and dEs.

After describing the data analysis and results, we compare our results with
some of the many hypotheses that have been proposed to explain the existence of
dEs or LSBs.  We also demonstrate that, although the objects were a potential
source of the missing mass in Coma, they probably play a negligible role in the
dynamics of the Coma cluster due to their small numbers.

In this paper, we assume a distance of 95 Mpc to Coma (H$_0$ = 75 km s$^{-1}$
Mpc$^{-1}$, v$_r$ = 7125 km s$^{-1}$, distance modulus = 34.89) and, therefore,
the scale is 0.46 kpc/arcsecond.

\section{{\sc OBSERVATIONS AND DATA REDUCTION}}

\subsection{{\em Observations}}

For full details regarding the observations and data reduction the reader is
referred to Bernstein et al. (1995).  Briefly, the Coma field was observed on
2--4~February~1991 from the KPNO 4-meter telescope, using a
backside-illuminated $1024\times1024$ Tektronix CCD at prime focus.  One pixel
spans 0.\arcsec473\ of the sky, giving an 8\arcmin\ field of view.  We made a
series of exposures in a typical ``shift and stare'' mode (Tyson 1986).  
The shifting reduced the high-S/N area of the final image to a 7.\arcmin5\
square.  Both R and B$_j$ filters were used, but the weather conditions were
much poorer for the B$_j$ exposures so those observations were not as deep as
those made with the R filter.  The FWHM of the point spread function in the
final R-band image is 1.\arcsec3 (see Figure 1 of Bernstein et al.).  The
field is centered at approximately $12^{\rm h}57^{\rm m}30^{\rm s}$,
$+28$\deg$09\arcmin 30\arcsec$ (equinox 1950), near the X-ray centroid of the
cluster (Ulmer, Wirth \& Kowalski 1992).  The giant elliptical (D) galaxies NGC~4874 and
NGC~4889 lie 40\arcsec\ and 280\arcsec\ off the NW and NE corners,
respectively, of our frame.

The data reduction specific to this paper was a three-step process.  In Step 1
(\S2.2), we used the FOCAS automated detection software (Valdes 1989) to find
all significant objects in the R image.  Step 2 (\S2.3) was to cull the low
surface brightness objects from the all-object FOCAS catalog.  This was done by
fitting gaussian profiles to all objects (in R band again), and selecting those
with fitted size and surface brightness beyond chosen thresholds.  Step 3
(\S2.4) was to derive true central surface brightnesses and scale lengths for
the selected low surface brightness objects in both R and B$_j$ images using
exponential profile fits, and to derive colors from common-aperture photometry. 
For Step 1 {\it only} it was necessary to remove the large diffuse-light
gradients from the R image.  Otherwise the object-finding algorithms would have
been overwhelmed and many objects would have been missed.  The diffuse-light
removal process and the FOCAS object-finding algorithms caused us to
underestimate the flux from extended low surface brightness objects, and
limited our completeness for very extended objects.  In Steps 2 and 3, however,
we used the {\it original} un-subtracted images for photometry; thus the
diffuse-light subtraction may have affected our completeness, but did not bias
the magnitudes and scales lengths which we later computed for the detected
objects.

\subsection{{\em Step 1: Removal of Diffuse Light and Identification of All
Objects}}
\label{diffuse}

The deep exposures we produced revealed diffuse emission that appears to be
primarily the result of an extended envelope from the D galaxies NGC~4874 and
NGC 4889.  The removal of the large-scale features from the R image began with
a fit of elliptical isophotes to the 23 brightest galaxies (and one bright
star).  The fitted ellipses were then subtracted from the image, allowing the
FOCAS software (Valdes 1989) to successfully search for dwarf galaxies where it
previously was ``blinded'' by flux gradients due to the emission from these
giant galaxies.  Regions where the elliptical isophotes were a poor fit were
masked and ignored in further processing.  Once these larger objects were
subtracted, we fit the large-scale gradients in the image by running a
15\arcsec-square median filter across the image.  These steps of galaxy fitting
and diffuse-light subtraction had to be iterated for best results, particularly
in the NW corner of the image, which contained steep diffuse-light gradients
from NGC~4874, another bright S0 galaxy, and a bright star.

FOCAS was then used to identify all objects within the frame, and calculate the
core magnitude ($m_c$), total magnitude ($m_t$), and position of all detected
objects.  We searched for LSBs using the FOCAS catalog as described below.  We
visually inspected the images as well, but we were not able to uncover any
objects missed by FOCAS.

The FOCAS detections are assuredly not spurious since we were able to detect,
in the companion B$_j$ image, all but one in the final set of LSBs initially
found in the R image.

\subsection{{\em Step 2:  Identification of Candidate Low Surface Brightness
Objects}}

Step 2 of the analysis was to take the objects found by FOCAS and select those
which were LSBs---i.e. those which were clearly resolved and which had low
central surface brightnesses.  Ideally this would have been done by fitting
exponential radial profiles to all objects, since LSBs and dEs normally have
such profiles (Bothun et al. 1989; Bothun et al. 1987).  We found, however, that
selecting on exponential-fit parameters produced a large number of spurious
candidates because our image is somewhat crowded, and neighbor objects often
confused the exponential fits.  We chose instead to select low surface
brightness objects on the basis of gaussian fits, which also give an indication
of the size and surface brightness of each object, but are less sensitive to
neighbors because the gaussian drops more rapidly.  Furthermore, many of the
objects are small, thus similar to the point spread function, which is nearly
gaussian.  The gaussian fits are better at distinguishing unresolved from
slightly resolved objects.

From the FOCAS R catalog we selected objects with $21.0<m_t<24.5$.  The FOCAS
catalog is 50\% complete for {\it unresolved} objects at $r=25.5$.  We selected
objects with $m_t-m_c<-0.5$ in order to limit stellar contamination.  For each
of the resultant 1400 objects we produced a radial integration of the surface
brightness on the {\it original} R image (before diffuse-light removal) and fit
a gaussian profile.  To choose LSB candidates, we needed a measure of the size
and of the surface brightness of each object.  For the size, we simply used the
gaussian $\sigma$.  The surface brightness is taken to be the mean surface
brightness inside an aperture whose radius is the point at which the object's
radial profile is 1.5 standard deviations above the local sky background.  The
size of the annulus used to measure the sky level and the standard deviation
was variable since the crowding was more severe in some parts of the image.

Figure 1a shows the distribution of all objects in the size versus 
surface-brightness plane.  We demanded that LSB candidates have
$\sigma>1.35\arcsec$ (above the horizontal line in Figure~1a), the level at
which we judged extended sources to be reliably discerned from the
$\sigma=0.56\arcsec$ (1.30\arcsec\ FWHM) point spread function.  LSB candidates
were also required to have average surface brightness (as defined above)
fainter than 27.5 R mag per square arcsecond (to the right of the dashed
vertical line).  This threshold was chosen to be about 0.2~mag brighter than
that of all extremely low surface brightness objects found by an {\em a priori}
visual inspection.  All candidate objects in this upper right quadrant were
then inspected visually; 18 were found to be true LSBs (diamonds in Figure~1a)
while the remainder were superpositions of 2 or more objects.  These 18 objects
are denoted with an L suffix (``lower'') in the Tables and Figures.

To correct for background contamination, we applied an identical selection
procedure to all objects in the 4 high-latitude control fields of
Bernstein et al (1995).  We found an average of only 2 LSBs per control field, so we
have henceforth assumed that nearly all of our detected LSBs are in fact Coma
cluster members.

We then extended our selection criteria to correspond roughly to the canonical
definition of an LSB as one having central surface brightness fainter than 23.5
B mag per square arcsecond.  For our gaussian R image fits, this corresponds to
26.6 mag per square arcsecond (the solid vertical line in Figure~1a).  Visual
inspection of the additional candidates to the right of the new surface
brightness threshold yielded, coincidentally, another 18 unblended objects.

To give an idea of how the gaussian selection criteria correspond to the more
usual exponential parameters, Figure~1b shows the scale length $\alpha$ and
central surface brightness $(\mu_0)_R$ of all 1400 objects, with the final 36
LSBs as diamonds.  All our LSBs meet the following two criteria:  central
surface brightness fainter than 24 R mag per square arcsecond, and scale length
$\alpha>0.9\arcsec$, which corresponds to about 0.4~kpc for the Coma distance
modulus of 34.89.  These are indicated by the solid lines in Figure~1b; note
that this is just a {\it description} of the sample, not the {\it definition}
of the sample, since we selected on the gaussian fit parameters.  Due to the
problems with diffuse light in Step 1, it is difficult to define a region over
which we can guarantee completeness of the sample.

\subsection{{\em Step 3:  Exponential Fits, Colors, and Total Magnitudes}}

The culling of Step 2 left us with a sample of 36 LSBs for analysis.  For
each of these objects, we integrated radial profiles on both the R and B$_j$
images---as in Step 2 we were using the {\it original} images, without
diffuse-light removal.  The centroids of the LSBs were fixed at the values
given by the centroid-finding algorithm in the IRAF package DAOPHOT.  These
agreed with the FOCAS centroids to within $\sim0.5\arcsec$, well below the
1.3\arcsec\ seeing FWHM, so sufficiently accurate for fitting radial profiles.

Next we fit exponential surface brightness laws to the radial profiles.  Fits
were performed to the linear fluxes (not magnitudes), and the sky level was
allowed to vary.  The object locations and the results of exponential fits to
the radial profiles are given in Table~1 for both R and B$_j$ data.  Figure~2
plots the R-band radial profile along with best exponential and gaussian fits
to each of the 36 LSBs.  We have retained the labels we initially gave the
objects, with the ``L'' suffix denoting the lower surface brightness half.  The
values of central surface brightness ($\mu_0$) are, of course, extrapolations
to $r=0$ because the seeing smooths the central sections.  Our LSBs are,
however, at least twice the size of the seeing disk (see Figure~1), so the
extrapolation is not too severe.  Examination of the Figure~2 profiles suggests
that central surface brightness is reliable at the level of a few tenths of a
magnitude.

In order to measure colors, we determined magnitudes inside a common {\it
aperture} for both the R and B$_j$ images of each LSB.  In Table~2 we list the
aperture magnitudes and the resultant color for each object.  Also given is the
radius used for the aperture, and the outer radius of the sky determination
annulus.  The inner radius of the sky annulus was the outer edge of the object
aperture.  Aperture sizes vary from object to object because of variable object
sizes and crowding.

We have calculated total magnitudes (listed in Table~1) by integrating the
exponential profiles to infinity.  Comparison of these total magnitudes and the
aperture magnitudes of Table~2 indicates that the uncertainties in the
magnitudes are typically less than 0.2~mag.

\subsection{{\em Exponential Fits, Central Surface Brightness, and
Total Magnitudes}}

In order to measure colors, we determined magnitudes using the same {\em fixed
aperture} for both R and B$_j$ for a given object.  The aperture radius
(``AR'') in Table 2 is not the size of the galaxy but the inner radius on the
local background annulus, and the outer radius (``SR'' in Table 2) is the outer
local background annulus.  The aperture radius is not the size of the LSBs, but
is the radius over which we performed the LSB integration prior to background
subtraction.  Since the field was crowded, we had to use different apertures
for each object, however.  The colors derived in this manner, and the
associated apertures for source and background, are given in Table 2 along with
the individually derived magnitudes.  We also calculated a total magnitude by
integrating the best fit exponential to infinity, and we used the best fits to
extrapolate inward to produce a central surface brightness (magnitudes/square
arcsecond).  We have included the resulting uncertainties in the estimates of
the central surface magnitudes given in Table 1.  Then, for self-consistent
(when considering exponential scale factors or central surface brightnesses)
estimates of the  total magnitude, the reader should use the magnitudes derived
from the exponential fits.  The results based on fixed apertures are to be used
for colors.

\subsection{\em{Interpretation of Objects as LSBs}}

From our analysis we conclude that in the final selection, the 36 objects are
LSBs, and that, based on comparison fields (cf. Bernstein et al. 1995), we estimate
that 90\% of our objects are located within the Coma cluster.  We re-examine
the conclusion that these objects are indeed LSBs in this subsection.  We begin
by noting that all the images were visually inspected and most of the objects
were well fitted with an exponential, and those few that were better fitted by
a gaussian had a gaussian $\sigma$ much larger than the seeing gaussian.  All
of the objects came from the initial FOCAS list of objects, and these images
were all visually inspected in order to reject blends.

As an alternative to the interpretation that the objects we found are LSBs, we
now demonstrate that the following interpretations are unlikely: (1) the
objects are globular clusters; (2) they are the blend of two point-like
sources; (3) they are a blend of two low surface brightness objects; and (4)
they are a blend of one low surface brightness object and one point-like
object.   We reject case ``1'' because the gaussian $\sigma$ of a point source
is $\sim 0.\arcsec6$ versus the observed value of $\simgreat 1.\arcsec4$ in our
sample (see Figure 1a), and at the distance of Coma, globular clusters will
appear as point sources (cf. Bernstein et al. 1995).  For case ``2'' the blending
requires a separation of about $1.\arcsec3$, and at such separations these
objects would be easily detected as blends.  For case ``3'' and case ``4'', we
require that such coincidences are to be within  about $1.\arcsec3$, but as the
total surface density of the objects is so small, the probability of such
chance occurrences is less than 1\%.  We conclude, therefore, that the large
majority of objects in our final selection are LSBs in the Coma cluster.

\section{{\sc DISCUSSION}}

\subsection{{\em Introduction}} 

As noted in the previous subsection, we have produced a list of objects of
which the large majority are LSBs in the Coma cluster.  For the sake of
simplicity, we will assume that {\em all} of these objects are LSBs in the Coma
cluster and will proceed with the discussion on this basis.  This sample is one
of extraordinarily faint, low surface brightness objects in a very rich
environment, and we begin by presenting 4 examples of these very faint objects
in Figure 3.

As a framework for our discussion, we consider just some (out of the myriad of
suggestions) of the hypotheses of the origin for (dwarf) LSBs: (1) they are the
result of fragmentation of larger galaxies caused by galaxy-galaxy interactions
in the cluster; (2) their number density is correlated with the spiral fraction
of the cluster (Turner et al. 1993); (3) they are related to the diffuse
extended envelopes of the D galaxies in this cluster which would make them
different from LSBs in clusters without D of cD galaxies; (4) they are objects
which have been confined by a hot intra-cluster medium and have faded from
their initial state; and (5) they are the result of large mass loss early in
their lifetime.  By relating to previous work, we show that none of these
scenarios is especially favored and that the origin of these objects remains a
mystery.  For general discussions of the origin and evolution of LSBs and dEs,
see Ferguson \& Binggeli (1994), Cole (1991), Bothun et al. (1989), and references
therein.

Our definition [$\alpha \simgreat 0.4$ kpc, $\mu_0 > $ 24.0\ (in R), mean
$\alpha \sim 0.6$ kpc, mean $\mu_0 \sim $ 26.0\ (R)] overlaps most closely with
the Turner et al. (1993) definition of {\em large} LSBs ($\alpha \sim 0.6$ to 1.5
kpc, $\mu_0$ $\sim 26.5$ in V), which most closely overlaps  with those 
of McGaugh \& Bothun (1994; mean values of $\alpha \sim 1.5$ kpc, $\mu_0
\sim 23.5$ in B) and Bothun et al. (1991; $\alpha \sim 0.7$ kpc,
$\mu_0 \sim 24.5$ in B). Turner et al. use the large LSB definition as
their main point of comparison with previous work.  For the sake of argument we
will use our definition below, but remind the reader that the standard set by
Bothun and co-workers over the past 10 years is for  B(0) to be fainter than 23
magnitudes per square arcsecond.  Also, for a discussion of the various types
of dE shapes and selection effects in clusters, see Bothun et al. (1989).

\subsection{{\em Comparison with Fornax and Virgo}}

To provide more details, as well as to help elucidate the sensitivity of our 
survey, we compare our results with previously cataloged objects.  This
comparison also demonstrates the plausibility of our assumption that these LSBs
are dwarf galaxies in the Coma cluster.  In Figure 4, we show how faint our
LSBs are with respect to the Fornax sample of Bothun et al. (1991).  When we
correct for cluster distance, we find that the brighter end overlaps the
Bothun et al. survey objects.  It is also informative to use Figure
4 to understand the sensitivity of the surveys as denoted by the curved lines. 
To the right of the left-most curved line is the region in which we would
expect to detect galaxies, given the angular diameter (smaller objects cannot
be distinguished from stars) and isophotal surface brightness limits (the
faintest level out to which an image is actually detected) noted next to these
curves.  These curves are derived by assuming an exponential form for the
galaxies of varying scale values as denoted by the diagonal dashed lines.  As
expected, most of our objects fall to the right of the left-most curved line,
based on the seeing (1.\arcsec3, FWHM) and the fluctuations in the background
in B$_j$ ($\sim 29^{th}$ magnitude per pixel).
 
Next, in Figure 5 we show our results in relation to those of Feguson \& Binggeli (1994).
Our sample is where we would expect to find it relative to the general
population of dEs, i.e., in the low surface brightness, low total brightness
portion of the plane, as shown in Figure 5.  The data in Figure 5 were compiled
from Kormendy (1985), Bothun et al. (1987), van der Kruit (1987),
Binggeli \& Cameron (1991, 1993), and Caldwell et al. (1992). 
From Figures 4 and 5, we infer that the properties of Coma LSBs overlap those
of LSBs or dEs in Fornax or Virgo in terms of their central surface
brightnesses and total magnitudes.

As an aside, we remark that, in Figure 5, it appears as though we have not been
able to detect larger objects (larger values of $\alpha$) for a fixed central 
surface brightness that would have populated the brighter side of the
luminosity distribution.  However, in Figure 4, it appears that we {\em did}
sample the brighter end of the distribution of LSBs (for a fixed central
surface brightness) quite well.  As noted previously, however, we do not
exclude the possibility that we have missed some larger objects (larger values
of $\alpha$), either by slipping below the FOCAS detection in Step 1, or because
the likelihood of these objects appearing as blends is larger than for those
with smaller values of $\alpha$, e.g., for a scale factor of $3\arcsec$, we
could, on average, integrate only about three to four scale lengths before
reaching the next object in the frame. But there is an inconsistency between
the interpretations of Figures 4 and 5 (one figure implies we may have missed
several faint extended objects for a fixed central surface brightness, the
other does not).  At least part of the discrepancy between conclusions about
our ability to find large-scale size LSBs, based on Figures 4 and 5, could be
due to discrepancies in distance estimates to the galaxies in the various
samples.  It is also possible that selection effects, or the manner in which
the total brightnesses (how far from the center a galaxy image was integrated
or what model was used) of the galaxies were calculated, could play a role in
producing discrepancies between different surveys.

Returning to the main point of this discussion, so far we have shown that the
LSBs in Coma have a size and brightness distribution which overlaps that of the
LSBs in other clusters [i.e., Fornax and Virgo; most of the Ferguson \& Binggeli (1994) dEs
are from the Binggeli \& Cameron (1991) survey of the Virgo cluster].

\subsection{{\em LSB Counts Versus Richness}}

When comparing the derived number density of LSBs with previous work, the
reader should be aware that the final results are very sensitive to definition,
e.g., by simply changing the range of acceptable scale factors for LSBs from
the range of 3.\arcsec5--9.\arcsec0 to all those $\geq$ 3.\arcsec0,
Turner et al. (1993) change the total number of LSBs in their sample by about a
factor of 4 (see also Figure 1 in this paper).  Furthermore, the range of
central surface brightnesses of a survey need also be considered, and we have
not corrected for any such effects.  For example, we only use those LSBs
actually detected in our analysis rather than those that extend over some
theoretically accessible magnitude range (e.g., Table 6, column 3 of
Turner et al.).  Therefore, although we find no evidence for trends in
number density in comparing our Coma data with the A3574 data of
Turner et al., one should keep in mind that the dependence of LSB density
on other cluster properties cannot be ruled out at this time, and that our
conclusions (below) are based only on a simplistic comparison (good to only a
factor of two) of the {\em large} LSB number counts of Turner et al. with
our LSB counts.  Furthermore, as we remark below, there is apparently a dearth
of brighter objects with large extents (values of $\alpha$ in the 10 to 30
range).  As noted by Turner et al., the difficulty in identifying
brighter, more extended, objects is crowding, such that it is not easy to
distinguish these objects from blends.  Thus, in comparing LSB number counts in
sparse versus crowded regions, it is also necessary to be aware that large
objects can be lost in the crowded fields, which can suppress the total number
counts in crowded fields.

With the above caveats in mind, we now consider the number counts of Coma 
versus A3574, as these clusters are widely different in richness.  Coma is
richness 2 and A3574 is richness 0 (Abell, Corwin \& Olowin 1989).  We can make this comparison
in two ways: first as a ratio of LSBs to ``ordinary galaxies'' (those brighter
than M$_B = -17$); and second, as a net surface density of LSBs.  Therefore,
effects due to galaxy-galaxy interaction might be observable.  We find that for
Coma and A3574 this ratio is about the same: 1.5/1 for Coma\footnote{Here we
have assumed an average B$_{j}-$R of 0.7 to convert the R luminosity function
for the core Coma (cf. Bernstein et al. 1995) to the B$_j$ band, and we have
converted our LSB surface density to LSB surface density per magnitude.} and
3.5/1 for A3574 (Turner et al. 1993).  Looking at this another way,  if the
fraction of LSBs were constant between the two rich clusters then we would
predict the number density for our sample to be about 16 times that of
Turner et al., whereas we find a factor of 8 increase ($\sim 800$/Mpc$^2$
versus $\sim 100$/Mpc$^2$ for A3574).  Although this suggests the existence of
a real effect, the accuracy of this comparison is only a factor of 2.  Also, if
we assume that the interaction rate depends mainly on the {\em square} of the
density of galaxies (in more massive/denser systems the velocities also tend to
be higher), we would expect a much stronger effect than the apparent factor of
2 effect seen here.  Therefore, although we cannot exclude the possibility, we
conclude that galaxy-galaxy interaction and fragmentation is not likely to be
the major cause of LSBs.

\subsection{{\em LSB Counts Versus Spiral Fraction}}

Related to the above hypothesis is the conjecture by Turner et al. (1993) that a
spiral fraction of 62\% might be related to the cause of the high density of
LSBs in A3574, as both the spiral fraction and LSB density are higher in A3574
than in Fornax (40\%, Godwin \& Peach 1982).  But Coma has a spiral fraction of
about 14\% (Dressler 1980), yet its surface density of LSBs is 8 times higher
than that for A3574, and the ratio of LSBs to ordinary galaxies is about the
same as for Fornax (1.5).  Thus, spiral fraction and LSB density do not seem to
be correlated.  Turner et al. note, however, that the situation is
complicated and that part of the difficulty may be related to the small angular
scale of our objects which prevents us from distinguishing between dEs and
dwarf Irregulars.
  
\subsection{{\em Relationship of LSBs to Extended Envelopes of D Galaxies}}

Next we consider the possibility that the LSBs in Coma are the result of the
fragmentation of extended envelopes around D or cD galaxies.  We cannot use a
comparison between Coma and A3574 as a discriminator, however, as both clusters
contain dominant galaxies.  Coma is classified as Bautz-Morgan II, and A3574 as
Bautz Morgan I (cf. Abell et al. 1989).  But, the similarity of the LSBs in
these clusters to those in Fornax fails to support the suggestion that the
diffuse envelope fragmentation is important for LSB formation.  For this model
of extended envelope fragmentation to work, the data also require an {\em ad
hoc} assumption that the process of the extended envelope fragmentation into
LSBs must scale with the ordinary galaxy density.  There is no direct evidence
to rule out such an extended envelope fragmentation hypothesis, however, until
more rich clusters are examined.

Another test of a possible LSB relationship to an extended envelope is to
compare the color of the diffuse envelope with the LSBs'.  From our own CCD
frames, we estimate a value of $1.2\pm0.3$ B$_{j}-$R for the color of the
diffuse emission.  The mean value of the color of the LSBs in our sample is
about 1.1, with a variance of measured values from $-0.7$ to about 2.  We also
searched for evidence of a color dependence of the LSBs on the distance from
the center of the D galaxies, but we found no statistically significant effect,
as can be seen in Figure 6a.  We also found that there is no obvious
(statistically significant) dependence of the density of LSBs on the distance
from NGC 4874.  Similarly, we found no statistically significant effect
relative to the distance from NGC 4889, the other D galaxy in the cluster.  We
conclude that there is no evidence that the presence of extended envelopes of D
or cD galaxies is related to LSBs, but we cannot exclude such a possibility.

We searched for such effects relative to the X-ray center as well.  Again we
found no statistically significant effects.  But as our frame only extends
about 1/2 of a core radius from the cluster center, we would not expect to see
any strong effects in these data.

\subsection{{\em LSB Color and Evolution}}

The color measurements can be used to search for other physical effects, and 
McGaugh \& Bothun (1994) explored the possibility (see also Bothun et al. 1991) that
most LSBs exist because they are ordinary galaxies that have faded to produce
low surface brightness systems.  In this scenario (which McGaugh \& Bothun did
not confirm in their data set), the fainter a galaxy is, the older, and hence
redder, it is.  Then, for the B$-$V color, McGaugh \& Bothun estimated a
reddening of 0.25 for every 1 magnitude decrease in central surface brightness. 
This estimate was based on standard stellar evolution models (Tinsley 1972). 
Since B$_{j}-$R is larger for the same stellar type, we would expect the effect
to be even more pronounced in the B$_{j}-$R versus central surface brightness
plane.  And we can see in Figure 6b that there could be $\sim 0.1-0.2$
magnitudes of reddening per 2 magnitudes of fading, but the data are also
consistent with no reddening.  Given the relatively large non-statistical
scatter in the data we cannot provide a valid (low chi-squared) best-fit
gradient to these data.  Combining our data with that of McGaugh \& Bothun,
however, we conclude that there is no evidence for color fading as the origin
of LSBs.  This conclusion is consistent with another test for such an effect
(also performed by McGaugh \& Bothun), i.e., color versus exponential scale,
which is shown in Figure 6c.

\subsection{{\em LSB M/L and Missing Mass}}

Another attribute that could be related to the origin of LSBs is their mass to
light ratio (M/L) and their relationship to the overall cluster mass.  We can
make an estimate of their masses by assuming that they have lasted the age of
the cluster.  Then the calculations of Bernstein et al. (1995) apply, and the M/L$_R$
we estimate has a lower bound\footnote{This is a lower bound in the sense that
a mass greater than this would produce stability.  This is, however, an upper
bound if considered in terms of the required lifetime of the galaxy in the
cluster.  For, as the age of the galaxy is reduced, so is its required (stable)
mass.} of about 60 M${\bigodot}$/L${\bigodot}$ which is consistent with dEs in
general (cf. Pryor 1992)\footnote{Here we have assumed that the color of a
typical elliptical galaxy is such that the difference between V and R is less
than 10\%.}.  We can estimate the total mass contribution to the cluster by
using the fraction of light that these objects contribute to the cluster as a
whole,  which is about 1/400 (cf. Bernstien et al.).  We use a value for the
optical luminosity for the cluster as a whole, of $10^{13}$ L${\bigodot}$
(Abell 1977).  Then, we find that the total contribution to the cluster
mass is a relatively small 10$^{12}$ M${\bigodot}$ [compared to the mass
required to virialize the cluster, i.e., about 10$^{15}$ M${\bigodot}$ (cf.
Sarazin 1986)], and even though LSBs dominate over ordinary large galaxies
in number, they do not contribute a dynamically significant amount of mass to
the cluster.  As LSBs are difficult to detect, they potentially could have been
a source for the missing mass in Coma, but this is not the case.  Even if we
had missed objects due to selection or crowding effects, these ``missed''
objects would have contributed to the diffuse light we detected in our image,
and as shown by Bernstein et al., there is not enough diffuse light to
account for the missing mass as long as M/L is less than several hundred.  Thus
the search for sources of the missing mass in Coma must continue.

\subsection{{\em LSB Formation Via Supernova Mass Ejection}}

Finally, we consider the hypothesis that the existence of dEs/LSBs in rich
clusters is due to large (90\%) mass loss due to supernova driven winds.  We
cannot directly test this hypothesis, but we can show that if dEs were
initially 10 times more massive and lost all this excessive mass to the
intra-cluster medium (ICM), then it is just possible to explain the presence of 
iron in the ICM (Henriksen \& Mushotsky 1986).  The argument goes as follows: an estimate
of gas in the cluster is about $10^{14}$ M${\bigodot}$ (White et al. 1993) over
approximately the same volume as the total light estimate of $10^{13}$
L${\bigodot}$ (i.e. about a 3 to 4 Mpc radius).  Then if the LSBs (or dEs) have
lost $10^{13}$ M${\bigodot}$ total to the ICM with solar abundance, this would
produce about 0.1 solar abundance in the cluster.  Within the accuracy of the
estimates and measurements (the iron abundance in the ICM of Coma is between
approximately 0.2 and 0.4 times solar, with 90\% confidence limits;
Henriksen \& Mushotsky), then, this is good agreement between hypothesis and
observation.  In this calculation, however, we have implicitly assumed that
most of the current mass of the dEs is baryonic, since we also assumed that 10
times the current mass has been ejected into the ICM, and that this ejected
mass was baryonic.  Thus, we deem this scenario unlikely, but we can neither
disprove it nor an alternative explanation of lower baryonic mass loss of
enriched material (i.e., higher amounts of iron and other heavy elements
relative to solar abundance).  A future line of investigation could be to
compare the iron abundance in the ICM with dE densities, but as most rich X-ray
bright clusters are farther away than Coma (cf. Sarazin 1986), this task
will be difficult to carry out.

\section{{\sc SUMMARY AND CONCLUSIONS}}

In summary, we have found 36 extremely low surface brightness dwarf galaxies in
the core of the Coma cluster.  We find {\em no} statistically significant
evidence to relate the origin or evolution of the LSBs to: fragmentation via
galaxy-galaxy interactions; the spiral fraction; the presence of D galaxies, as
the Coma LSBs seem typical of LSBs found in other clusters that do not contain
D galaxies; or, long-lived color fading.  There is also no statistically
significant dependence on color with respect to the projected distances of the
LSBs from key locations such as the D galaxies or the X-ray center. 
Furthermore, these LSBs probably do not contribute significantly to the overall
mass of the cluster or to the iron in the hot intra-cluster  medium.  Thus, the
origin and evolution of LSBs and their relationship to rich clusters remain
unclear.  Observations of other rich clusters and observations farther from the
Coma cluster core are necessary to help solve the mystery.  Finally, although
numerous, LSBs (or faint objects in general) that were missed in previous
surveys of Coma cannot make up for the missing dynamical mass in the cluster.

\section{{\sc ACKNOWLEDGEMENTS}}

We thank M. Bell for her programming support.  Thanks to P. Teague for
assistance in the Coma observing run, and to P. Guhathakurta for allowing us to
commandeer images from other projects for use as control fields.  GMB was
generously supported by AT\&T Bell Laboratories and the Bok Fellowship from
Steward Observatory during the tortuously extended duration of this work. 
Ulmer thanks A.~Sandage for discussions at the inception of this project, and
Northwestern University and NASA for partial support.  We thank the referees,
K. O'Neil and G. Bothun, for useful and insightful comments.  We also thank A.
Ulmer for providing useful comments.

\renewcommand{\arraystretch}{1.2} 
\begin{table}
\begin{center} {\sc Table 1: LSBs' Positions and Exponential Parameters}~\\ 
\begin{tabular}{lccrcrrcccccc} 
\sc
\\[-0.1in]
\hline 
\hline 
id\# & h & m & s & d & $'$ & $''$ & m$_R$ & m$_{B_j}$ & S$_R$ & S$_{B_j}$ & 
$\alpha_R$ & $\alpha_{B_j}$\\
\hline
18 & 12 & 57 & 3.46 & 28 & 6 & 11.0 & 21.50 & 22.66 & 23.82 & 25.10 & 1.06
& 1.24\\
18L & 12 & 57 & 4.53 & 28 & 6 & 8.6 & 23.69 & 25.11 & 25.66 & 26.94 & 0.92
& 0.88\\
17 & 12 & 57 & 5.47 & 28 & 10 & 2.6 & 21.66 & 23.67 & 23.88 & 25.38 & 1.25 
& 1.45\\
16 & 12 & 57 & 5.90 & 28 & 7 & 37.2 & 21.39 & 22.38 & 23.69 & 24.75 & 1.26
& 1.30\\
17L & 12 & 57 & 7.02 & 28 & 8 & 39.3 & 23.52 & 25.29 & 26.40 & 27.19 & 1.95
& 1.34\\
16L & 12 & 57 & 7.45 & 28 & 11 & 49.1 & 25.17 & 26.66 & 25.97 & 29.58 & 0.85
& 1.71\\
15L & 12 & 57 & 8.17 & 28 & 11 & 11.6 & 23.41 & 25.01 & 25.93 & 26.71 & 1.46
& 1.14\\
14L & 12 & 57 & 8.46 & 28 & 6 & 20.8 & 23.41 & 24.48 & 25.61 & 26.63 & 1.36
& 1.13\\
15 & 12 & 57 & 8.68 & 28 & 6 & 5.0 & 22.15 & 23.95 & 23.88 & 26.41 & 0.96
& 1.51\\
13L & 12 & 57 & 8.96 & 28 & 8 & 38.3 & 24.10 & 24.73 & 25.65 & 27.08 & 0.93
& 1.38\\
14 & 12 & 57 & 9.43 & 28 & 11 & 16.6 & 21.75 & 22.77 & 24.09 & 25.37 & 1.27
& 1.45\\
12L & 12 & 57 & 10.58 & 28 & 8 & 57.0 & 23.38 & 24.48 & 25.97 & 26.69 & 1.50
& 0.99\\
13 & 12 & 57 & 10.87 & 28 & 11 & 51.4 & 21.83 & 23.20 & 24.10 & 25.32 & 1.25
& 1.27\\
12 & 12 & 57 & 11.38 & 28 & 6 & 49.5 & 22.23 & 24.14 & 23.81 & 26.42 & 0.80
& 1.47\\
11 & 12 & 57 & 12.85 & 28 & 12 & 7.1 & 20.46 & 21.49 & 23.80 & 25.03 & 2.04
& 2.37\\
11L & 12 & 57 & 15.66 & 28 & 10 & 28.7 & 22.51 & 23.54 & 25.31 & 26.81 & 1.50
& 2.34\\
10 & 12 & 57 & 16.16 & 28 & 10 & 4.8 & 21.49 & 22.38 & 24.44 & 25.71 & 1.75
& 2.10\\
9 & 12 & 57 & 17.10 & 28 & 10 & 51.9 & 21.15 & 22.04 & 23.09 & 24.51 & 1.04
& 1.44\\
\hline
\end{tabular} ~\\ 
\end{center}

\end{table}

\renewcommand{\arraystretch}{1.2} 
\begin{table}
\begin{center} {\sc Table 1 (cont.)}~\\ 
\begin{tabular}{lccrcrrcccccc} 
\sc
\\[-0.1in]
\hline 
\hline 
id\# & h & m & s & d & $'$ & $''$ & m$_R$ & m$_{B_j}$ & S$_R$ & S$_{B_j}$ & 
$\alpha_R$ & $\alpha_{B_j}$\\
\hline 
10L & 12 & 57 & 18.00 & 28 & 10 & 30.1 & 23.62 & 24.64 & 25.71 & 26.74 & 1.05
& 0.52\\
8 & 12 & 57 & 19.87 & 28 & 5 & 58.0 & 21.92 & 22.78 & 23.93 & 25.25 & 1.19
& 1.47\\
7 & 12 & 57 & 21.24 & 28 & 6 & 29.2 & 22.51 & 23.42 & 24.49 & 25.81 & 1.22
& 1.28\\
9L & 12 & 57 & 21.60 & 28 & 11 & 36.2 & 23.34 & 24.53 & 25.14 & 26.76 & 0.98
& 1.24\\
6 & 12 & 57 & 22.75 & 28 & 7 & 17.0 & 21.39 & 22.73 & 23.50 & 24.87 & 1.26
& 1.42\\
8L & 12 & 57 & 23.58 & 28 & 5 & 51.5 & 22.40 & 23.66 & 25.10 & 26.20 & 1.47
& 1.27\\
7L & 12 & 57 & 24.08 & 28 & 8 & 51.3 & 22.68 & 23.87 & 24.58 & 25.87 & 1.07
& 1.17\\
5 & 12 & 57 & 25.20 & 28 & 8 & 52.5 & 22.08 & 22.83 & 23.76 & 25.27 & 0.82
& 1.30\\
6L & 12 & 57 & 25.70 & 28 & 10 & 4.6 & 22.40 & 23.33 & 25.63 & 26.83 & 1.84
& 2.02\\
5L & 12 & 57 & 27.21 & 28 & 9 & 28.3 & 22.83 & 24.82 & 26.02 & 26.96 & 1.85
& 1.38\\
4 & 12 & 57 & 27.86 & 28 & 9 & 35.6 & 21.38 & 22.40 & 23.70 & 24.98 & 1.37
& 1.54\\
3 & 12 & 57 & 27.97 & 28 & 11 & 29.3 & 22.83 & 23.32 & 24.58 & 25.84 & 1.25
& 1.49\\
2 & 12 & 57 & 29.31 & 28 & 12 & 25.2 & 22.21 & 22.71 & 23.67 & 24.91 & 0.71
& 1.25\\
4L & 12 & 57 & 29.95 & 28 & 8 & 49.2 & 22.62 & 23.69 & 24.89 & 26.12 & 1.33
& 1.33\\
3L & 12 & 57 & 30.35 & 28 & 7 & 33.8 & 23.19 & 24.41 & 25.30 & 26.62 & 1.24
& 1.23\\
2L & 12 & 57 & 33.66 & 28 & 10 & 59.8 & 23.01 & 23.95 & 24.89 & 26.45 & 0.97
& 1.49\\
1L & 12 & 57 & 35.93 & 28 & 5 & 45.4 & 23.48 & 24.91 & 25.49 & 27.29 & 1.03
& 1.39\\
1 & 12 & 57 & 36.29 & 28 & 9 & 5.7 & 22.27 & 23.37 & 23.87 & 25.31 & 0.87
& 1.15\\
\hline 

\end{tabular} ~\\ 
\end{center}
\end{table}
\clearpage
\noindent
\underline{Notes to Table 1} \\ Column 1: identification number; Columns 2--7:
RA and Dec in 1950 equinox; Columns 8 and 9: total magnitudes in R and B$_j$
based on integrating the best fit exponential model to infinity for the R and
B$_j$ images; Columns 10 and 11: the central surface brightness (also denoted
$\mu_0$ elsewhere) for R and $B_j$; and, Columns 12 and 13: the exponential
angular scale factor in arcseconds for R and B$_j$.  The estimated
uncertainties in the central surface brightness and scale factors increase from
the brighter to the fainter objects as follows: the average uncertainty in the
angular scale of the LSBs is 0.03 to 0.05 (R) and 0.04 to 0.06 (B$_j$) 
arcseconds, and for the central surface brightnesses 0.06--0.08 (R) and
0.04--0.06 (B$_j$).  The estimated uncertainties in total magnitudes are
typically less than 0.2 (see text).
         
\begin{table}
\begin{center}{\sc Table 2: Fixed Aperture Magnitudes and Colors}~\\
\begin{tabular}{lccccc}
\sc
\\[-0.1in]
\hline
\hline
id\# & m$_R$ & m$_{B_j}$ & $B_{j}-R$ & AR ($''$) & SR ($''$)\\
\hline
18 &21.35&22.52&1.17&4.73&7.57\\
18L&23.54&25.39&1.85&4.07&6.15\\
17 &21.53&22.56&1.03&4.49&7.57\\
16 &21.51&22.40&0.89&4.73&7.57\\
17L&23.69&24.87&1.17&4.54&6.15\\
16L&24.50&27.93&3.43&3.07&5.91\\
15L&23.60&25.20&1.60&4.02&5.91\\
14L&23.21&24.33&1.12&4.02&6.15\\
15 &22.01&23.87&1.86&5.20&7.57\\
13L&24.00&24.67&0.66&3.55&4.73\\
14 &21.84&22.94&1.10&4.97&7.57\\
12L&23.69&24.39&0.70&4.73&5.91\\
13 &22.09&23.10&1.01&4.73&7.57\\
12 &22.17&23.77&1.60&5.68&8.04\\
11 &20.45&21.47&1.02&7.80&9.22\\
11L&22.40&23.47&1.07&4.82&6.15\\
10 &21.45&22.45&1.00&7.80&9.22\\
 9 &20.93&22.06&1.13&6.39&8.04\\
\hline
\end{tabular}

\end{center}

\end{table}

\begin{table}
\begin{center}{\sc Table 2 (cont.)}~\\
\begin{tabular}{lccccc}
\sc
\\[-0.1in]
\hline
\hline
id\# & m$_R$ & m$_{B_j}$ & $B_{j} - R$ & AR ($''$) & SR ($''$)\\
\hline
10L&23.67&25.77&2.11&4.78&5.91\\
 8 &21.65&22.72&1.07&4.97&7.09\\
 7 &22.47&23.86&1.39&5.20&7.57\\
 9L&23.27&24.50&1.23&5.16&7.09\\
 6 &21.33&22.46&1.13&3.55&5.44\\
 8L&22.37&23.77&1.40&5.44&7.57\\
 7L&22.49&23.85&1.37&4.49&7.57\\
 5 &21.95&22.91&0.96&5.91&7.33\\
 6L&22.36&23.50&1.15&7.76&8.99\\
 5L&23.20&25.47&2.27&7.09&8.51\\
 4 &21.37&22.53&1.15&7.57&8.99\\
 3 &22.11&23.31&1.20&4.49&7.57\\
 2 &22.06&22.77&0.71&3.78&7.57\\
 4L&22.51&23.79&1.28&4.49&7.09\\
 3L&23.11&24.67&1.56&4.59&7.57\\
 2L&22.96&23.66&0.71&5.34&7.57\\
 1L&23.37&25.57&2.20&4.49&7.57\\
 1 &22.17&23.27&1.10&4.73&7.57\\
\hline

\end{tabular}

\end{center}
\end{table}

\clearpage \noindent \underline{Notes to Table 2} \\ Column 5 is the radius in
arcseconds of the aperture used for photometry.  Sky was determined in an
annulus from this radius to the ``sky radius'' in column 6.  The uncertainties
in total magnitudes and colors are dominated by the choice of the background,
and scatter in the background, rather than by counting statistics.  The errors
in color correspond to less than 0.1 magnitudes for the brighter objects, and
increase up to about 0.5 magnitudes for the fainter objects in the sample.  The
estimated uncertainties in total magnitudes are typically less than 0.2 (see
text).


\newpage
\section{{\sc FIGURE CAPTIONS}}

Fig. 1 \_\_ (1a): average surface brightness versus gaussian best fit $\sigma$
= ``radius''; diamonds are the LSBs culled from examining all objects above the
horizontal line.  (1b): central surface brightness versus exponential scale
factor; the diamonds are the same objects as shown in 1a.  The vertical lines
demark the initial (dashed) and final (solid) selection limit of our sample
(see text).

Fig. 2 \_\_ Thirty-six separate plots showing the radial intensity profiles of
the LSBs (R images only) we found, along with both the gaussian (dotted curves)
and exponential best fits.  The right hand corner of each plot contains the ID
numbers related to Tables 1 and 2.  The bottom X-axes are in kpc, the top are
in arcseconds.  The best fit local background has been subtracted from these
data.

Fig. 3 \_\_  Images of the sample LSBs from our survey.  Best fit local
background has been subtracted.  The designation of these objects in Tables 1
and 2 are, reading clockwise from the top left corner, 2L, 4L, 13L and 7L.  The
spatial scale on these plots is arcseconds.  The surface brightness scale
given below each plot is in R magnitudes per square arcsecond.  The pixel size
is 0.\arcsec473.

Fig. 4 \_\_ The total brightness in blue versus the central surface brightness. 
The crosses are from Bothun et al. (1991).  The asterisks are our data and
assume H$_0$ = 75 km/sec-Mpc.  The open diamonds are our sample corrected to
the distance of Fornax.  The solid curves show the selection function that
relates the limiting central surface brightness, and the limiting diameter of
the object, to the exponential scale factor given by $B_{tot}$= $-.6689 +5.0
\times \log[(\mu_{lim}-\mu_0)/\theta_{lim}] + \mu_0$, where $\mu_{lim}$ is the
limiting isophotal magnitude, $\theta_{lim}$ is the limiting diameter in arc
seconds, and $\mu_0$ is the observed central surface brightness (see also
Bothun et al.; Gilmore, King \& van der Kruit 1989, page 259).  The arrows show the
direction of motion on the central surface brightness total apparent magnitude
plane based on a correction for distance.

Fig. 5 \_\_ The absolute surface brightness versus absolute blue magnitude for
various classes of galaxies, shown with plusses, are from Ferguson \& Binggeli (1994).  The
open diamonds are our LSBs.

Fig. 6 \_\_ Top (6a) color versus distance from NGC 4874.  Center (6b): average
LSB color (from Table 2) versus central surface brightness in the blue (from
Table 1).  Bottom (6c): color versus exponential scale factor in arcseconds. 
In all of these plots LSB 16L has been excluded due to the faintness, and hence
large uncertainty, in the B$_j$ value.
   


\newpage

\begin{figure}[tbh]
\centerline
{\psfig{file=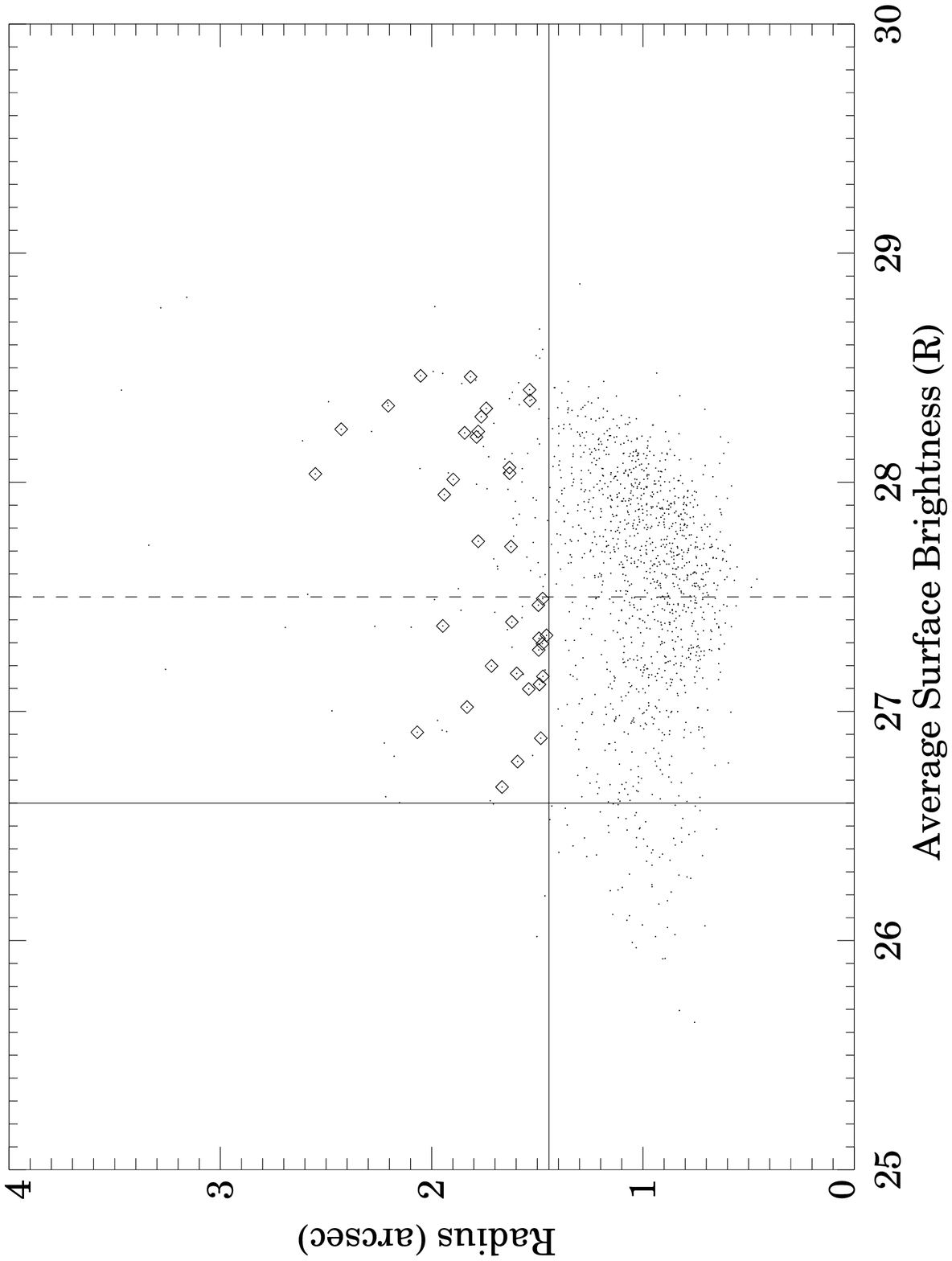,width=6.5in,angle=180}}
\label{Figure 1a}
Figure 1a~\\
\end{figure}

\begin{figure}[tbh]
\centerline
{\psfig{file=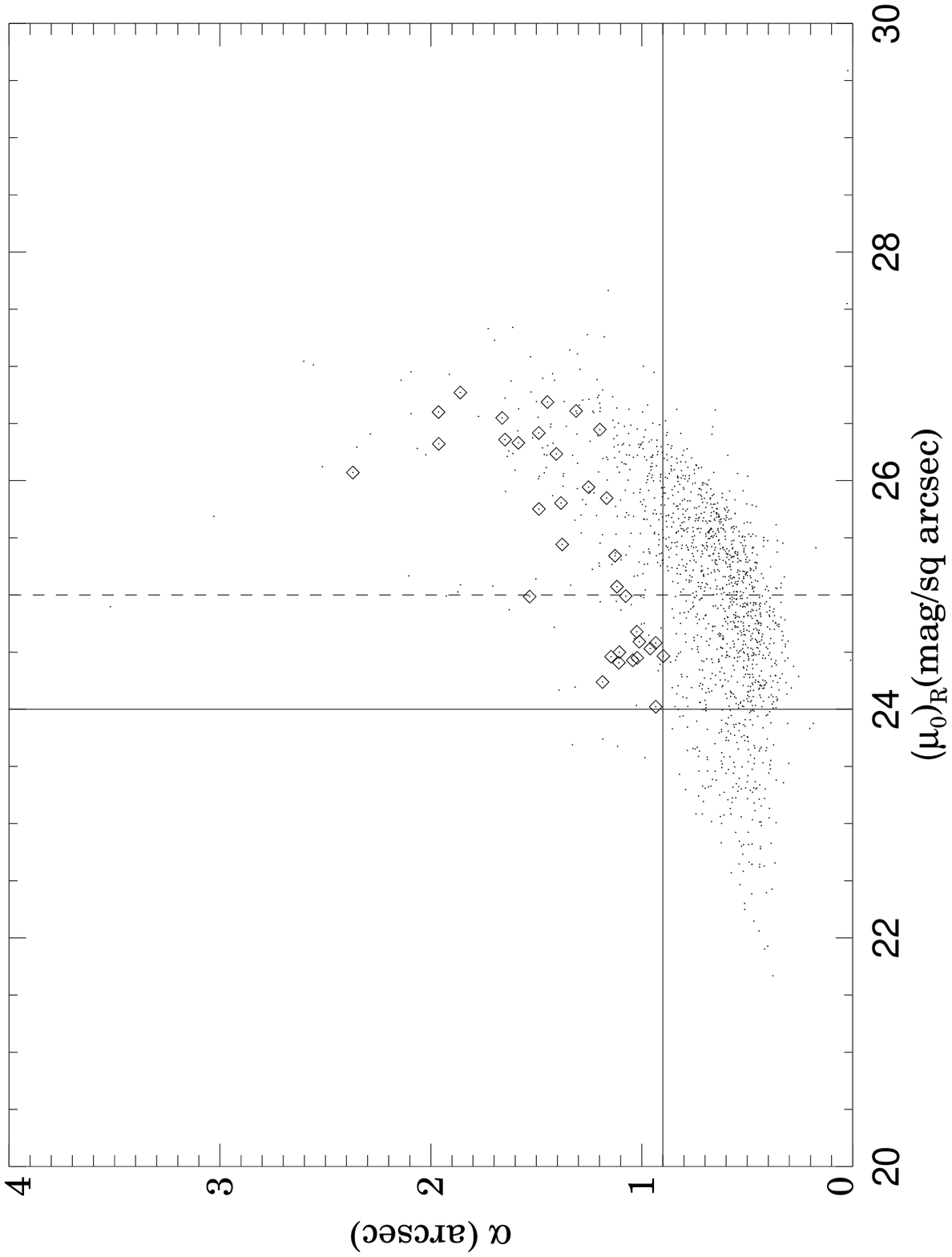,width=6.5in,angle=180}}
\label{Figure1b }
Figure 1b
\end{figure}

\begin{figure}[tbh]
\centerline
{\psfig{file=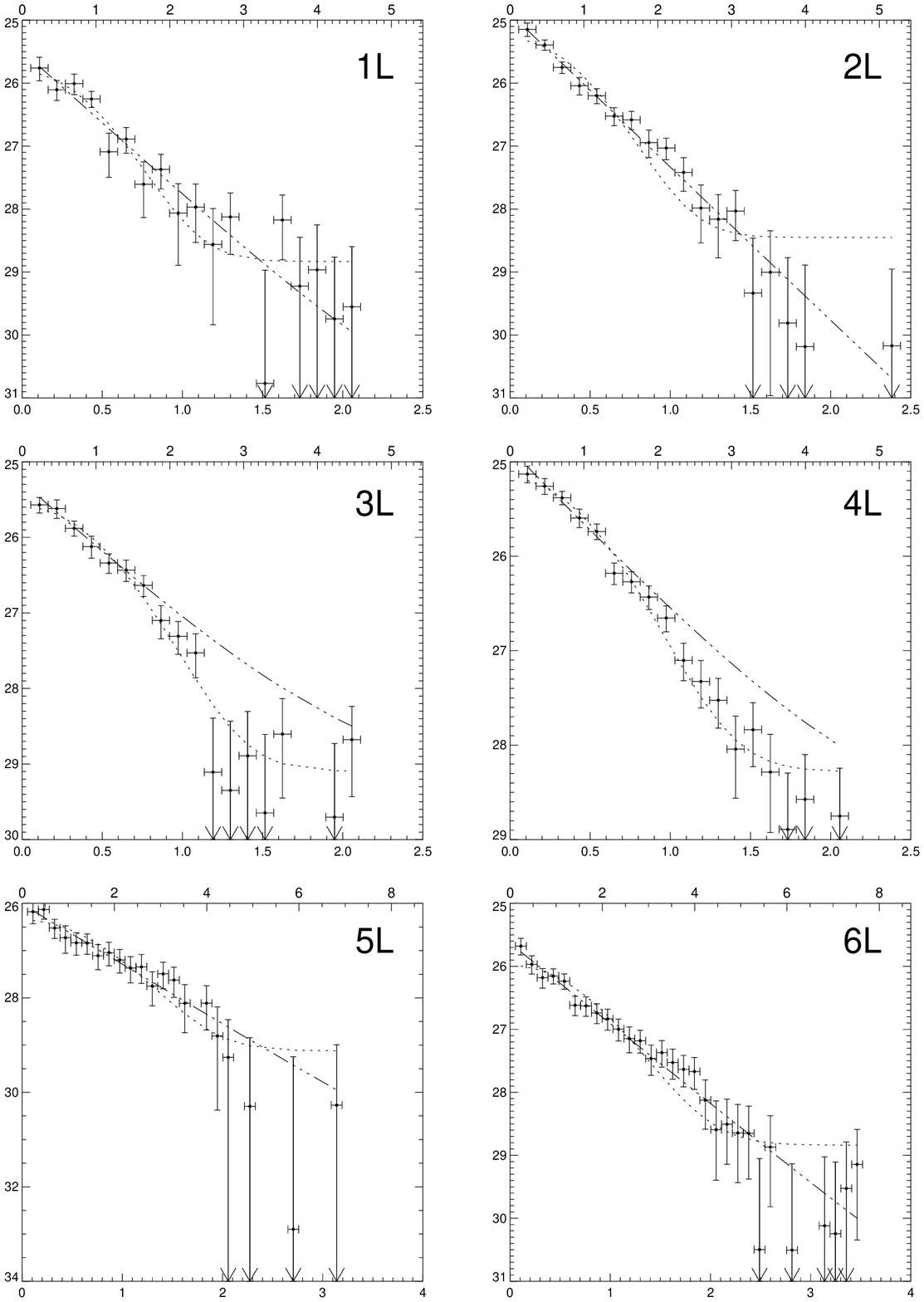,width=6.5in,angle=0}}
\label{Figure 2a}
Figure 2~\\
\end{figure}

\begin{figure}[tbh]
\centerline
{\psfig{file=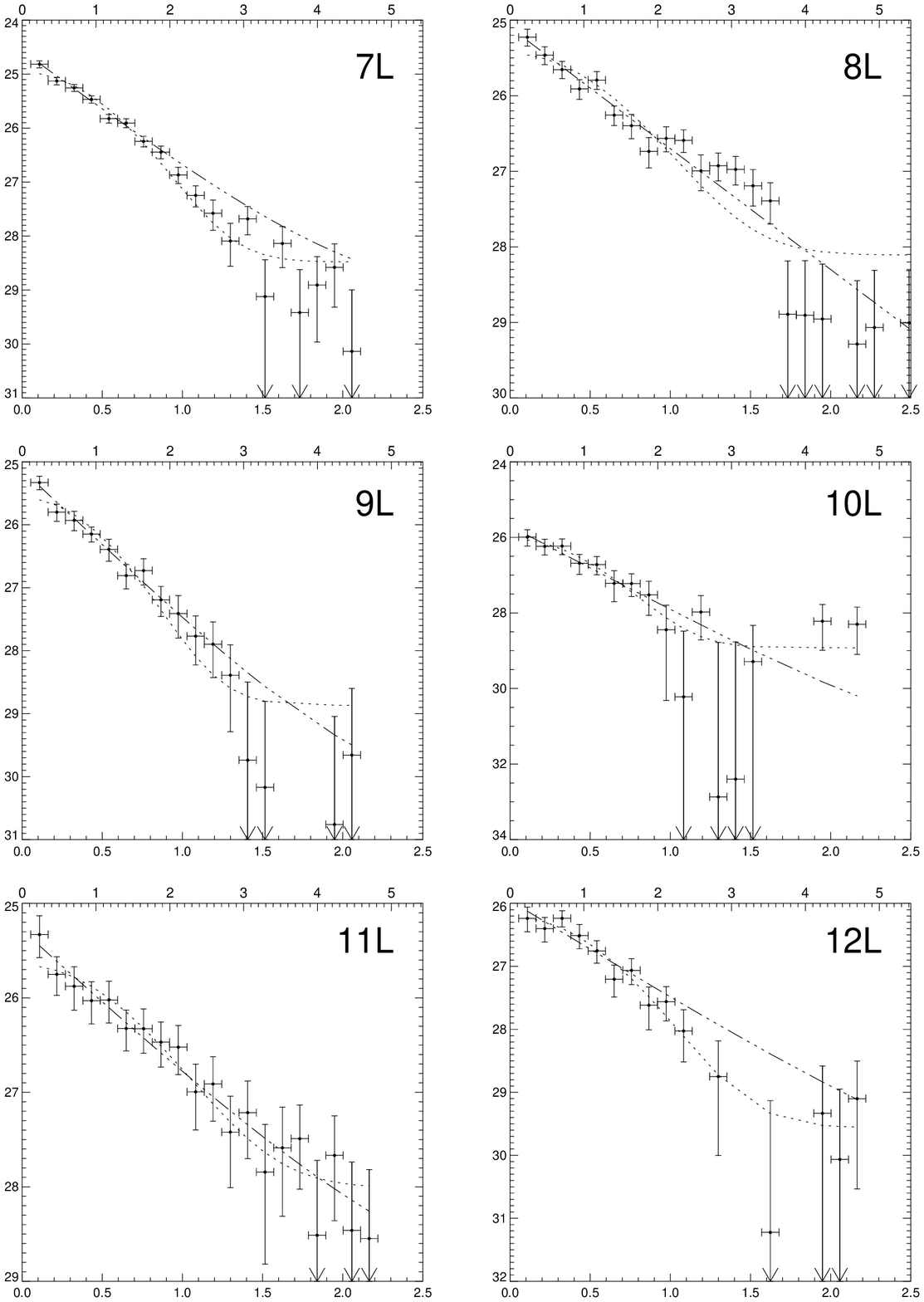,width=6.5in,angle=0}}
\label{Figure 2b }
Figure 2 (cont)~\\
\end{figure}

\begin{figure}[tbh]
\centerline
{\psfig{file=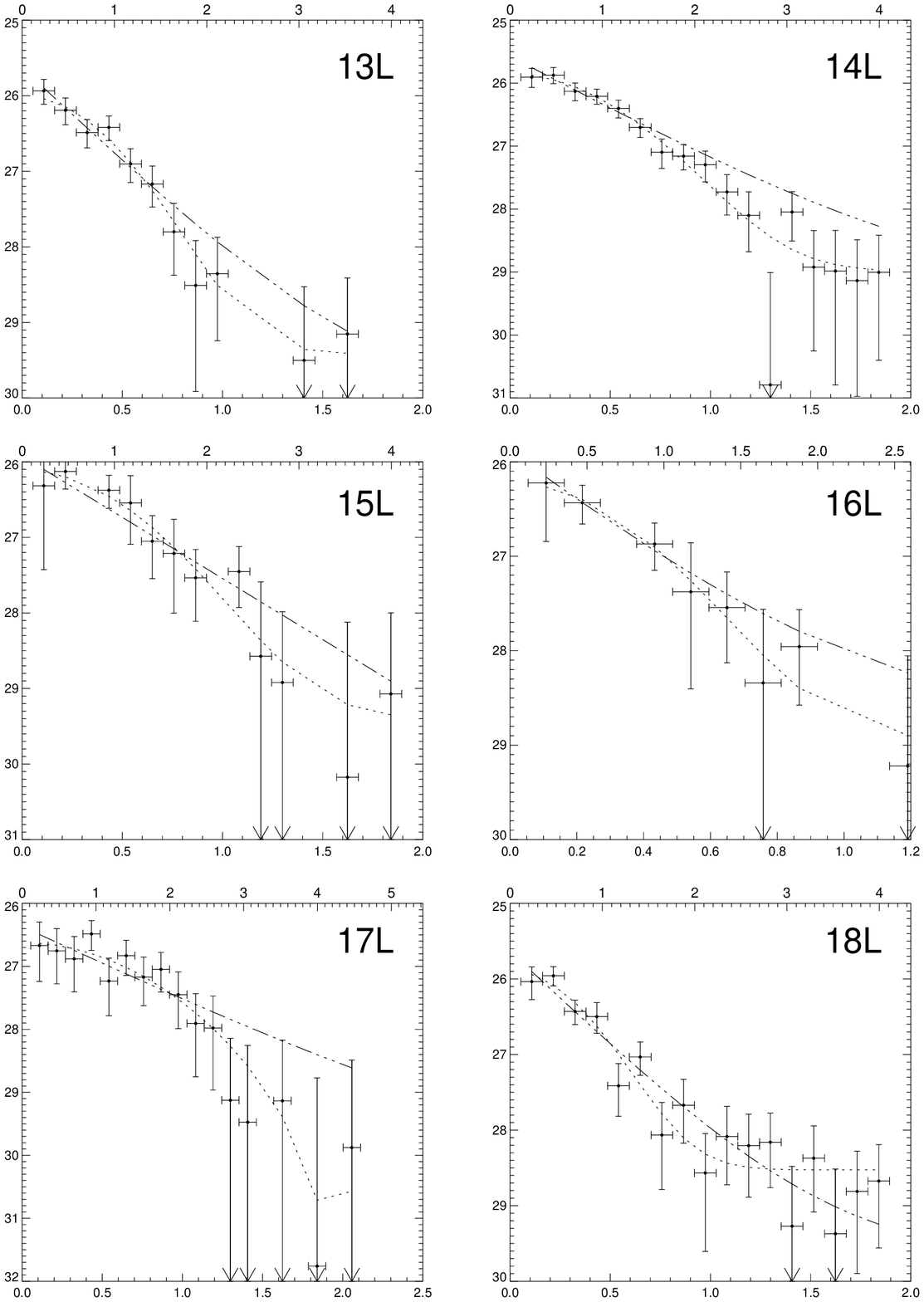,width=6.5in,angle=0}}
\label{Figure 2c }
Figure 2 (cont)~\\
\end{figure}

\begin{figure}[tbh]
\centerline
{\psfig{file=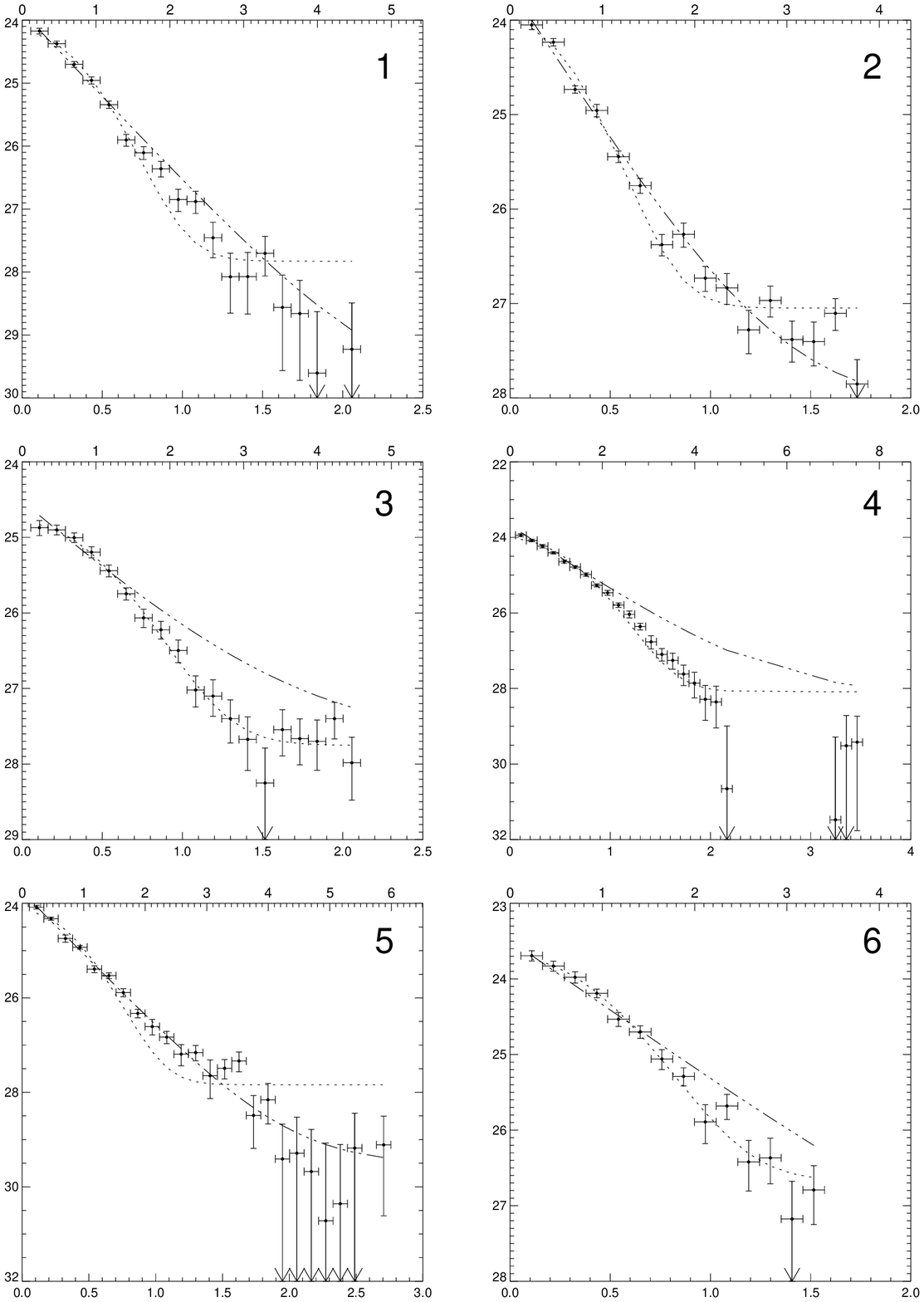,width=6.5in,angle=0}}
\label{Figure 2d}
Figure 2 (cont)~\\
\end{figure}

\begin{figure}[tbh]
\centerline
{\psfig{file=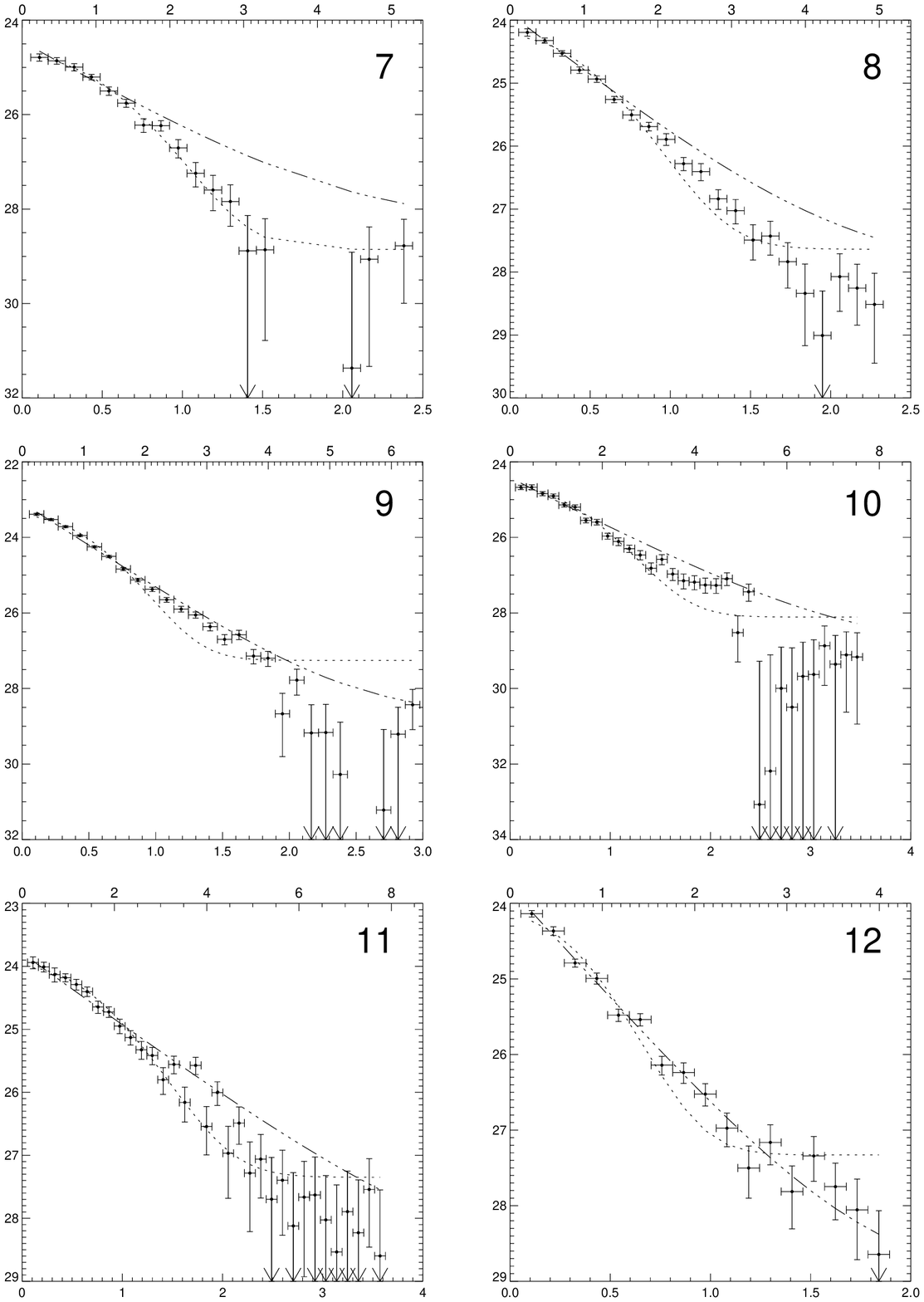,width=6.5in,angle=0}}
\label{Figure 2e}
Figure 2 (cont)~\\
\end{figure}

\begin{figure}[tbh]
\centerline
{\psfig{file=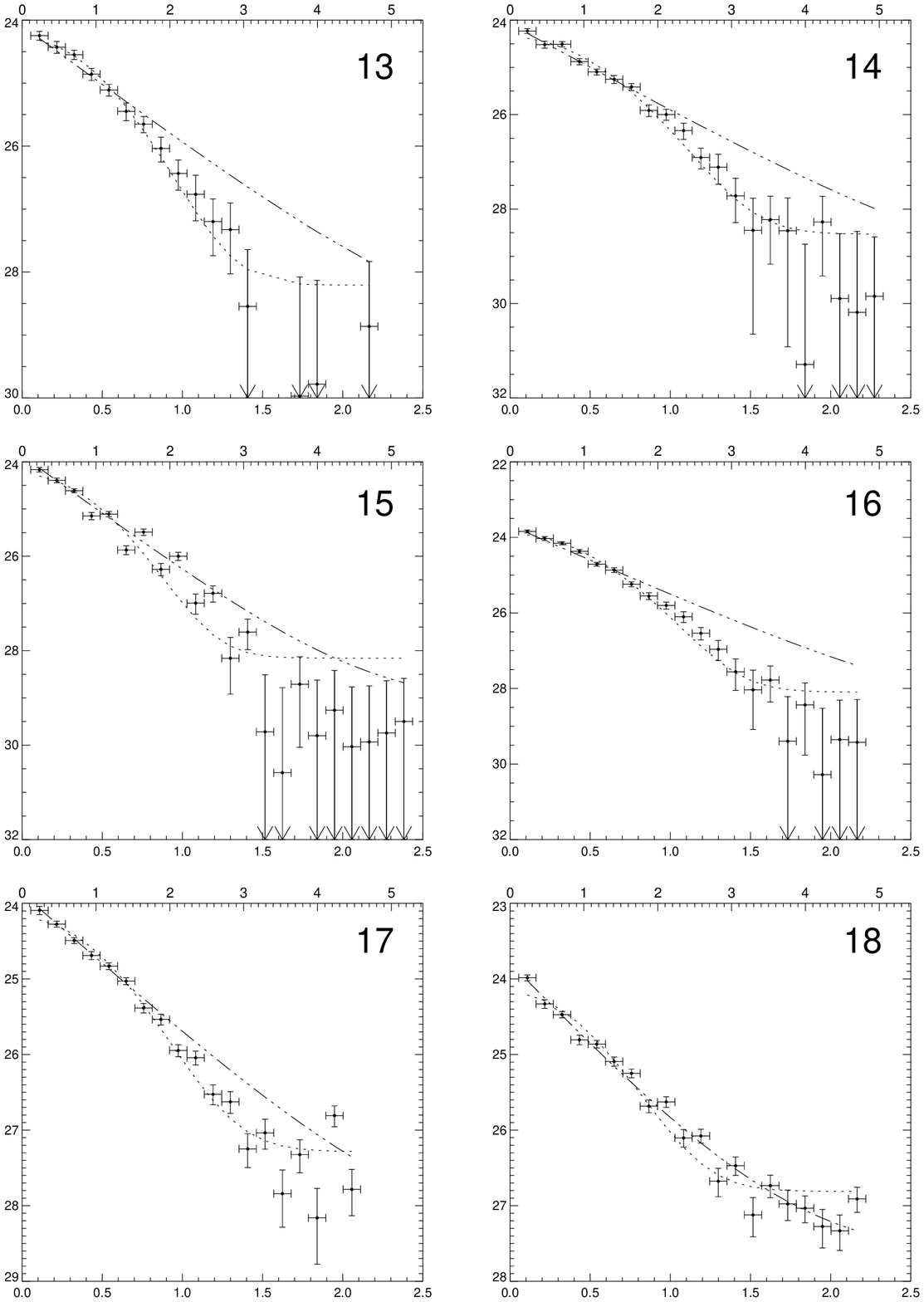,width=6.5in,angle=0}}
\label{Figure 2f}
Figure 2 (cont)~\\
\end{figure}

\begin{figure}[tbh]
\centerline
{\psfig{file=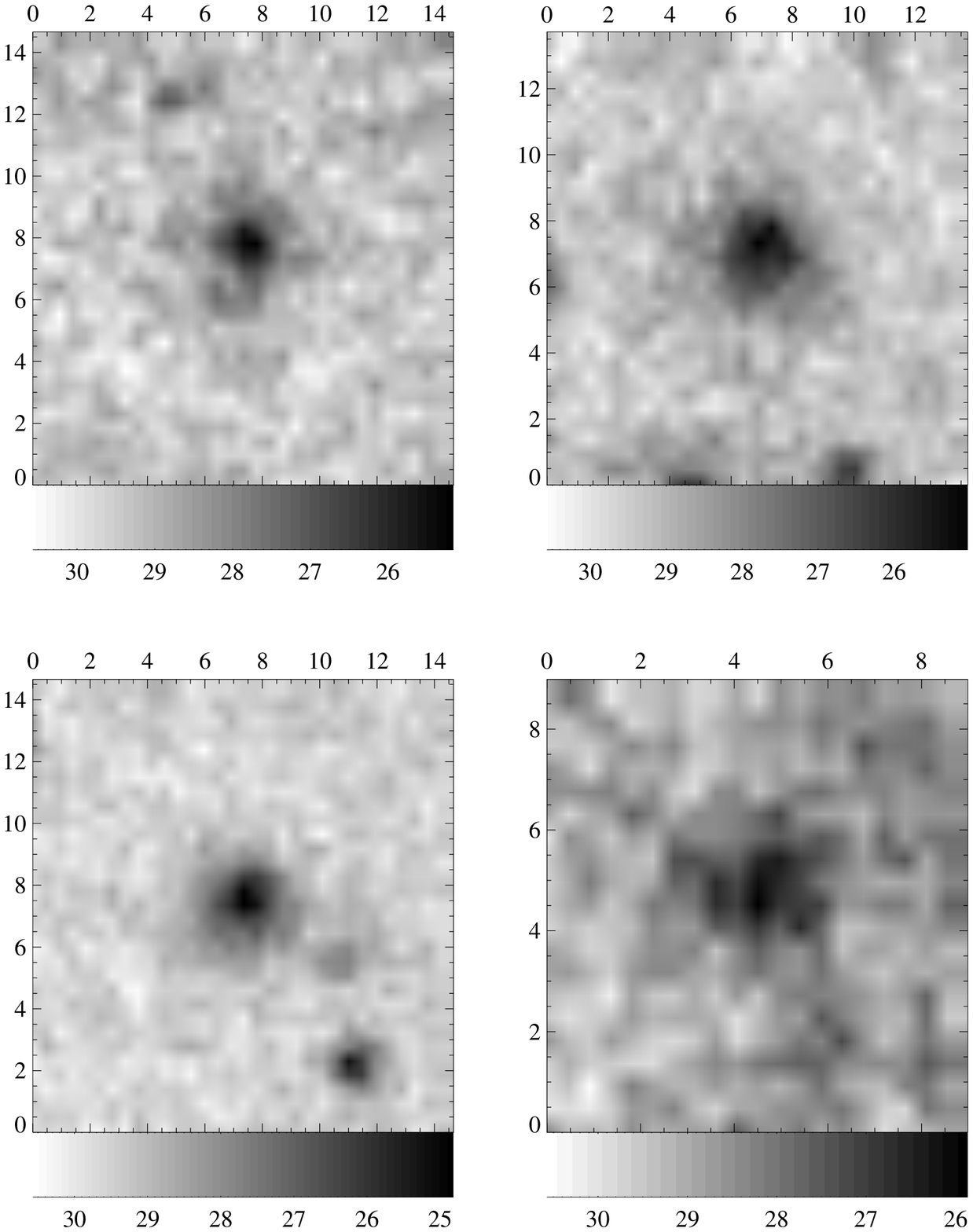,width=6.5in}}
\label{Figure 3}
Figure 3~\\
\end{figure}

\begin{figure}[tbh]
\centerline
{\psfig{file=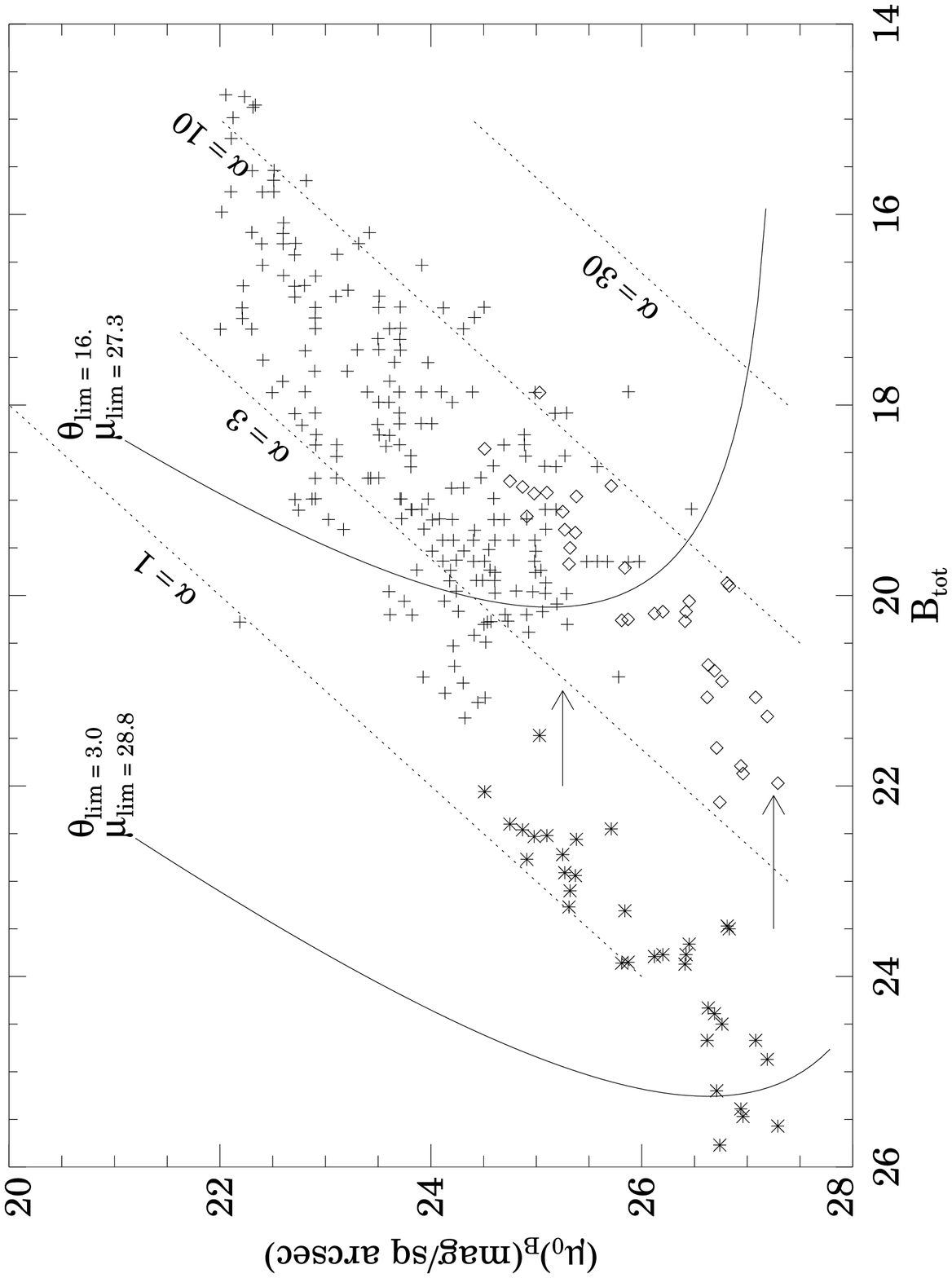,width=6.5in,angle=180}}
\label{Figure 4}
Figure 4~\\
\end{figure}

\begin{figure}[tbh]
\centerline
{\psfig{file=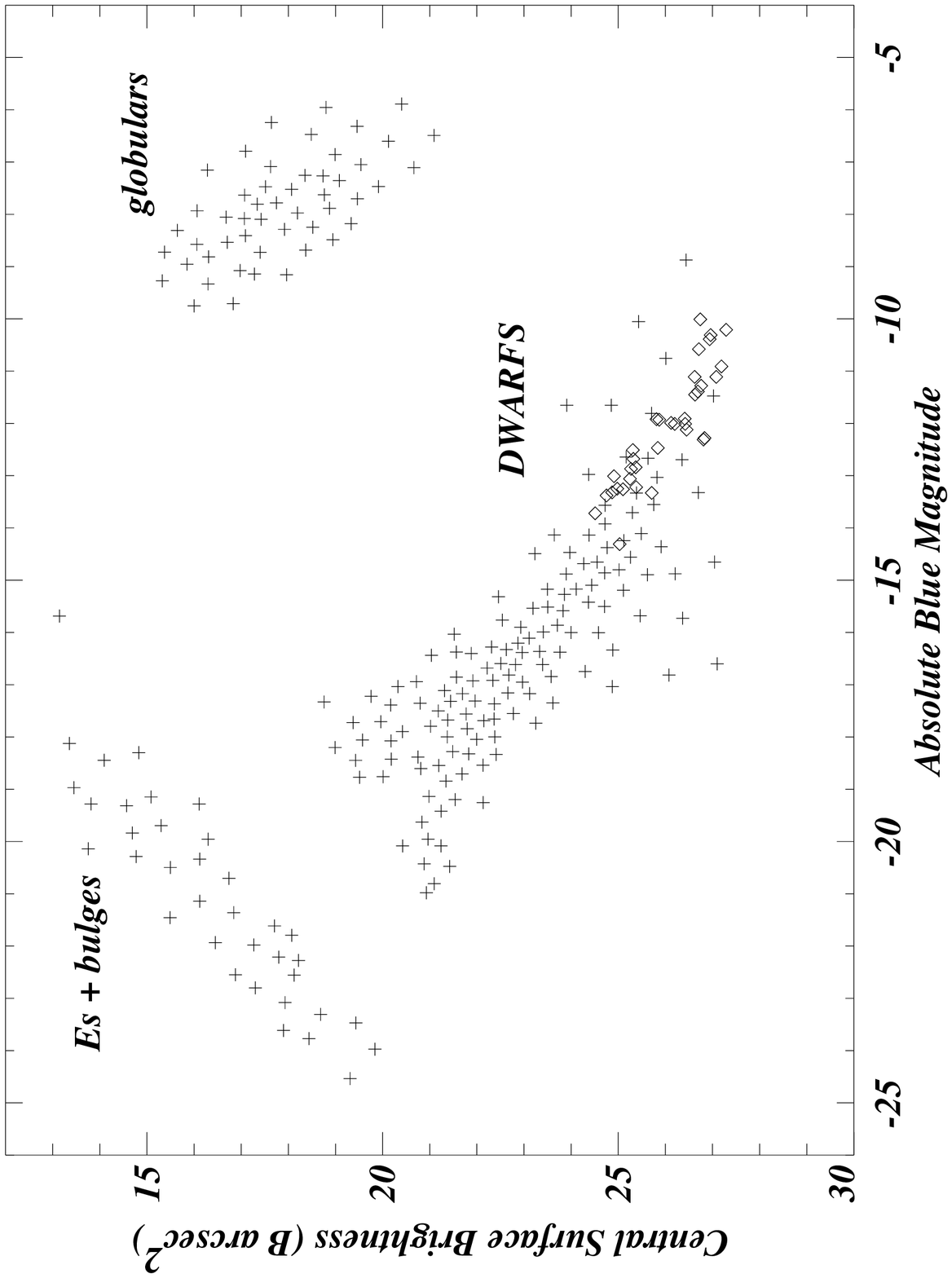,width=6.5in,angle=180}}
\label{Figure 5}
Figure 5~\\
\end{figure}

\begin{figure}[tbh]
\centerline
{\psfig{file=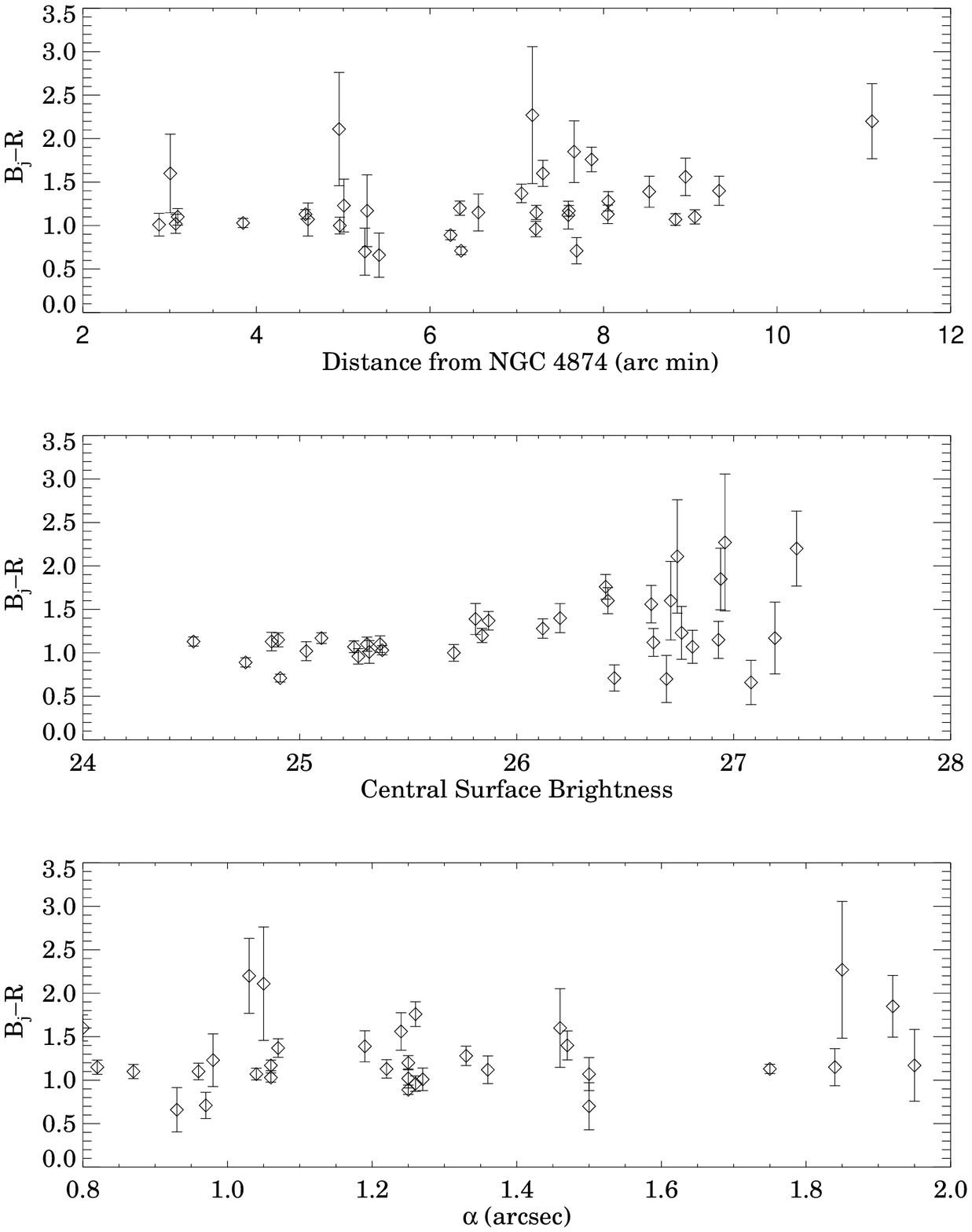,width=6.5in}}
\label{Figure 6}
Figure 6~\\
\end{figure}


\end{document}